\documentstyle[12pt]{article}
\renewcommand
\baselinestretch{1.3}

\def\beq{\begin{equation}}
\def\brr{\begin{array}}
\def\err{\end{array}}
\def\eeq{\end{equation}}
\def\bea{\begin{eqnarray}}
\def\eea{\end{eqnarray}}
\def\bs{\bigskip}
\def\tr{\mbox{Tr}\, }

\def\wt{\widetilde}
\def\wh{\widehat}

\def\nn{\nonumber}

\begin{document}


\hfill November 1993

\vspace*{3mm}

\begin{center}

{\LARGE \bf
One-loop renormalization of higher-derivative 2D dilaton gravity}

\vspace{4mm}

\renewcommand
\baselinestretch{0.8}
\medskip

{\sc E. Elizalde}
\footnote{E-mail: eli @ ebubecm1.bitnet} \\
Department E.C.M. and I.F.A.E., Faculty of Physics,
University of  Barcelona \\  Diagonal 647, 08028 Barcelona
 \\
and Blanes Center for Advanced Studies, CSIC, 17300 Blanes, Spain
\\
{\sc S. Naftulin} \\
 Institute for Single Crystals, 60 Lenin Ave., 310141
Kharkov, Ukraine \\
and Torselin, Sci. and Manufacture Centre, 310105 Kharkov, Ukraine
\\
{\sc S.D. Odintsov}\footnote{
Also at Tomsk Pedagogical Institute, 634041 Tomsk, Russian
Federation.
E-mail: odintsov @ ebubecm1.bitnet} \\
Department E.C.M., Faculty of Physics,
University of  Barcelona \\ Diagonal 647, 08028 Barcelona,
Spain \\

\renewcommand
\baselinestretch{1.4}

\vspace{5mm}

{\bf Abstract}
\end{center}

A theory of higher-derivative 2D dilaton gravity which has its
roots in
the massive higher-spin mode dynamics of string theory is
suggested. The
divergences of the effective action to one-loop are calculated,
both in
the covariant and in the conformal gauge. Some technical problems
which
appear in the calculations are discussed. An interpretation of the
theory as a particular D=2 higher-derivative $\sigma$-model is
given.
For a specific case of higher-derivative 2D dilaton gravity, which
is
one loop multiplicatively renormalizable, static configurations
corresponding to black holes are shown to exist.

\vspace{4mm}

\newpage

\noindent 1. {\bf Introduction}.
It is well known by now that the one-dimensional string (or the
one-dimensional $\sigma$-model) describes the so-called 2D gravity.
Such a theory has been very popular recently as a toy model for the
study of formal questions of quantum gravity, for the investigation
of the black hole structure and Hawking radiation, for its
interesting connections with conformal field theory, and so on. A
huge volume of literature on this field exists already and it is
generally  expected that the study of 2D gravity will help us in
the construction of a consistent theory of 4D quantum gravity.

Different models of 2D gravity have been considered. A very popular
one, which follows from a one-dimensional $\sigma$-model, is
described by the action
\beq
S= - \int d^2x \, \sqrt{g}\,  \left[ c_1(\Phi) g^{\mu\nu}
\partial_\mu
\Phi  \partial_\nu \Phi + c_2 (\Phi) R + V(\Phi) \right],
\label{1}
\eeq
where $\Phi$ is the dilaton field. However, this action is actually
the one-dimensional analog of the action which describes the
massless modes of the string. In order to take into account the
first level massive higher-spin modes, one has to modify the
standard bosonic $\sigma$-model, including also all possible terms
with quartic derivatives \cite{1}. Different approaches have been
developed so far \cite{1,2} for the description of the massive
string excitations, but only in the linear field approach.

It seems quite reasonable to formulate the theory of 2D gravity
which stems from the massive mode string dynamics and to study this
theory along the same lines as the usual 2D dilaton gravity. Of
course such a theory will be one with higher derivatives. Its
investigation may be useful in order to understand general
properties of quantum gravity, and also perhaps for
obtaining  the dynamics of massive string modes, since it
seems that this 2D gravity with higher derivatives should be easier
to understand than real strings.

Motivated by these considerations, we start from the following
action which includes all possible quartic derivative terms:
\bea
S &=& - \int d^2x \, \sqrt{g}\,  \left[ a_1(\Phi) g^{\mu\nu}
g^{\alpha
\beta} \partial_\mu \Phi  \partial_\nu \Phi  \partial_\alpha \Phi
\partial_\beta \Phi +  a_2(\Phi) g^{\mu\nu} g^{\alpha \beta}
\partial_\mu \Phi  \partial_\alpha \Phi  \nabla_\nu  \partial_\beta
\Phi \right. \nn \\ & +&
 a_3(\Phi) g^{\mu\nu} g^{\alpha \beta} \nabla_\mu
\partial_\alpha \Phi  \nabla_\nu  \partial_\beta \Phi +  a_4(\Phi)
g^{\mu\nu}  \partial_\mu \Phi  \partial_\nu \Phi  \Box \Phi +
a_5(\Phi) g^{\mu\nu} \Box \partial_\mu \Phi  \partial_\nu \Phi  \nn
\\ & +&  a_6(\Phi)   \Box \Phi \Box \Phi +  a_7(\Phi) \Box^2  \Phi
+   a_8(\Phi) g^{\mu\nu} \epsilon^{\alpha\beta}\nabla_\mu
\partial_\alpha \Phi \nabla_\nu \partial_\beta \Phi \nn \\
&+&  a_9(\Phi) \epsilon^{\mu\nu} \Box \partial_\mu \Phi
\partial_\nu \Phi +   a_{10}(\Phi) \epsilon^{\mu\nu}
g^{\alpha\beta} \partial_\mu \Phi  \partial_\alpha \Phi  \nabla_\nu
\partial_\beta \Phi +  a_{11}(\Phi) R  g^{\mu\nu}  \partial_\mu
\Phi \partial_\nu \Phi \nn \\
&+&  a_{12}(\Phi)   g^{\mu\nu}  \partial_\mu R \partial_\nu \Phi
+   a_{13}(\Phi) R \Box \Phi +   a_{14}(\Phi)  \Box R +
a_{15}(\Phi) R^2 +  a_{16}(\Phi) \epsilon^{\mu\nu}  \partial_\mu R
\partial_\nu \Phi \nn \\
&+& \left.  C_1(\Phi) g^{\mu\nu}  \partial_\mu \Phi  \partial_\nu
\Phi  +   C_2(\Phi) R   +   C_3(\Phi) \Box \Phi + V(\Phi) \right].
\label{2}
\eea
Here we suppose that all the functions of the dilaton $\Phi$
(coefficients) are analytic, $\epsilon^{\mu\nu}$ is an
antisymmetric tensor, the dimensions are: $[a] =L^2$,  $[C]
=L^0$,   $[V] =L^{-2}$, and the minus sign in front of the action
is chosen for convenience.

Integrating the action (\ref{2}) by parts and dropping total
derivatives (in full analogy with string theory \cite{1}), we
obtain
\bea
S &=& - \int d^2x \, \sqrt{g}\,  \left[ Z_1(\Phi) g^{\mu\nu}
g^{\alpha
\beta} \partial_\mu \Phi  \partial_\nu \Phi  \partial_\alpha \Phi
\partial_\beta \Phi +  Z_2(\Phi) g^{\mu\nu} g^{\alpha \beta}
\partial_\mu \Phi  \partial_\alpha \Phi  \nabla_\nu  \partial_\beta
\Phi \right. \nn \\ & +&
 Z_3(\Phi) g^{\mu\nu} g^{\alpha \beta} \nabla_\mu
\partial_\alpha \Phi  \nabla_\nu  \partial_\beta \Phi +  Z_4(\Phi)
R g^{\mu\nu}  \partial_\mu \Phi  \partial_\nu \Phi   +
Z_5(\Phi) R \Box  \Phi  + Z_6(\Phi) R^2  \nn \\
&+& \left.  C_1(\Phi) g^{\mu\nu}  \partial_\mu \Phi  \partial_\nu
\Phi  +   C_2(\Phi) R  + V(\Phi) \right].
\label{3}
\eea
In the next sections we will discuss the quantum structure of the
action (\ref{3}), which is going to be our starting point. In
particular,
we will calculate the one-loop divergences of the theory given by
(\ref{3}) in covariant and conformal gauges.
\bs

\noindent 2. {\bf One-loop renormalization}.
Let us start the calculation of the one-loop divergences of the
theory under discussion. In what follows we shall assume that
$Z_6(\Phi)$ has no zeros, and that
\beq
\Delta \equiv \det \left( \brr{cc} 4 Z_6 & -Z_5 \\ -Z_5 & Z_3 \err
\right) \neq 0.
\label{4}
\eeq
Such restriction appears in the course of the calculation, but its
precise physical meaning is not clear. We are going to work in the
background field method, according to which we set
\beq
\Phi \longrightarrow \Phi + \varphi, \ \ \ \ \ g_{\mu\nu}
\longrightarrow  g_{\mu\nu} + h_{\mu\nu},
\eeq
where $\varphi$ and $ h_{\mu\nu}$ are quantum fields, and the
following notations will be used: $h = g^{\mu\nu}  h_{\mu\nu}$ and
$ \bar{h}_{\mu\nu}= h_{\mu\nu} - \frac{1}{2} g_{\mu\nu} h$. Working
in the covariant effective action formalism ---developed for 2D
dilaton gravity (\ref{1}) in Refs. \cite{3}--- the problem reduces
to the calculation of the tr log from the fourth order
differential operator $\wh{H}$, which is essentially the second
functional derivative of the action (\ref{3}).

This operator contains $3\times 3$-matrices acting on the space of
quantum fields $\Phi^i \equiv \{ \varphi, h, \bar{h}_{\mu\nu} \}$.
If minimal gauge conditions are used, it takes the form
\beq
\wh{H}_{ij} =  \wh{K}_{ij} \Box^2 + \wh{L}_{ij}^{\mu\nu\lambda}
\nabla_\mu \nabla_\nu \nabla_\lambda +   \wh{M}_{ij}^{\mu\nu}
\nabla_\mu \nabla_\nu +   \wh{U}_{ij}^{\lambda}  \nabla_\lambda
+ \wh{Y}_{ij}.
\label{5}
\eeq
Without loss of generality, we take $ \wh{L}^{\mu\nu\lambda}$ and
$\wh{M}^{\mu\nu}$ to be fully symmetric in their greek indices.

As the matrices $\wh{K}$,  $\wh{L}$ and $\wh{M}$ do not commute
with the covariant derivative, a naive choice of their components
can get changed when integrating by parts. The remedy to this
situation is found in 't Hooft and Veltman's procedure \cite{4},
which yields unique Hermitian matrices, according to the rule
\bea
\wh{K} & \longrightarrow & \wh{K}' = \frac{1}{2} (\wh{K} +
\wh{K}^T),  \nn \\
 \wh{L}^{\mu\nu\lambda}  & \longrightarrow &
\wh{L}^{'\mu\nu\lambda}  = \frac{1}{2} (  \wh{L}^{\mu\nu\lambda}
-
  \wh{L}^{T\mu\nu\lambda}) + \frac{2}{3} (g^{\nu\lambda} \nabla^\mu
\wh{K}^T + g^{\mu\lambda} \nabla^\nu \wh{K}^T + g^{\mu\nu}
\nabla^\lambda \wh{K}^T  ), \nn \\
 \wh{M}^{\mu\nu}  & \longrightarrow &   \wh{M}^{'\mu\nu}  =
\frac{1}{2} (  \wh{M}^{\mu\nu}  +  \wh{M}^{T\mu\nu}) - \frac{3}{2}
\nabla_\lambda  \wh{L}^{T\mu\nu\lambda}+ \nabla^\mu  \nabla^\nu
\wh{K}^T  \nn \\ && \ \ \ \ \ +  \nabla^\nu  \nabla^\mu \wh{K}^T +
g^{\mu\nu} \Box \wh{K}^T,
\label{6}
\eea
and so on. Notice that the matrices $ \wh{U}^\lambda$ and $\wh{Y}$
have no effect on the one-loop divergences.

The divergent part of
the $\tr \log \wh{H}$ (modulo surface terms) may be expressed as
follows
\bea
\frac{i}{2} \left. \tr \log \wh{H} \right|_{div}  &=& \frac{i}{2}
\tr \log \left[ \wh{1} \Box^2 + \wh{S}^{\mu\nu\lambda}  \nabla_\mu
\nabla_\nu \nabla_\lambda  + \wh{N}^{\mu\nu}  \nabla_\mu \nabla_\nu
+ \wh{U}^{\lambda}  \nabla_\lambda + \wh{Y}\right]_{div}  \nn \\
&=& \frac{1}{32\epsilon}  \int d^2x \, \sqrt{g}\,  \left[ 2 \tr (
\wh{S}^{\mu\nu\lambda}  \wh{S}_{\mu\nu\lambda} ) + 3 \tr   (
\wh{S}^{\lambda}  \wh{S}_{\lambda} ) - 16 \tr \wh{N}_\nu^\nu
\right],
\label{7}
\eea
where $\wh{S}^{\mu\nu\lambda}$ and $ \wh{N}^{\mu\nu}$ are assumed
to be completely symetric in their greek indices, $\wh{S}^\lambda
\equiv \wh{S}_\nu^{\ \nu\lambda}$, and $\epsilon =2\pi (n-2)$. So
far
we have assumed that a minimal gauge of the standard type exists,
what is not evident in the case under discussion. Expanding the
action (\ref{3}) in powers of the quantum fields, one verifies that
there appear no terms of the form $\bar{h} \Box^2 \bar{h}$, so that
the matrix $\wh{K}$ in Eq. (\ref{5}) is degenerate. (In higher
dimensions, the corresponding term in standard $R^2$-gravity comes
from the Weyl tensor squared, which vanishes identically for
$d=2$). Possible solutions of this problem are the following. (i)
One may gauge the field $\bar{h}_{\mu\nu}$ away by adopting the
conformal gauge and by working in this conformal gauge. (ii) One
may instead
invent some procedure in order to make the matrix $\wh{K}$ in
(\ref{5}) non-degenerate ``by hand'', for instance, by adding some
term which should not influence the divergences. (iii) And one may also
consider a (non-standard) non-linear gauge, for example,
including curvature terms in the gauge condition.
In what follows we will apply the first two procedures and will
show that both give equivalent off-shell expressions for the
 one-loop divergences (up to surface terms).

In order to modify at the quantum level the second variation of the
action, let us consider the following term
\beq
\delta S = - \frac{\xi}{2}  \int d^2x \, \sqrt{g}\,   Z_6(\Phi)
\bar{h}^{\mu\nu} \Box [2\wt{R}_{\mu\nu}-\wt{R}  \wt{g}_{\mu\nu} ].
\label{8}
\eeq
Here the field $\Phi$ is classical, while the quantities with
tildes contain both background and quantum components, and hence
must be Taylor expanded (to first order in fluctuations). The
weight factor $\xi$ is arbitrary. As the expression in the square
brackets on the r.h.s. of Eq. (\ref{8}) is zero precisely when $d=2$,
the divergent part of the effective action should not depend on
$\xi$. (Sometimes a non-essential renormalization of the background
field is compulsory in order to eliminate the $\xi$-dependence
\cite{5}).

Adding the term (\ref{8}) to the initial action (as has been proposed
for the 2D dilaton gravity (\ref{1}) in Refs. \cite{5,6}) may actually
change the structure of the operator $\wh{H}$ (\ref{7}).
Finally, at the end of the calculations one can put $\xi$ equal to
zero. However, since the intermediate expressions acquire a pole at
$\xi =-1$ or $\xi =0$, we will consider the region $\xi >0$ only.
Note also that it is not evident that the ghost operator will
become both minimal and non-degenerate for $\xi \neq 0$, and this
should be checked directly for the gauge under discussion.

Let us choose the gauge-fixing action in the form
\beq
 S_{GF} = -   \int d^2x \, \sqrt{g}\,  \chi^\mu C_{\mu\nu}
\chi_\nu,
\label{9}
\eeq
where
\bea
\chi^\mu &=& - \nabla_\nu \bar{h}^{\mu\nu} + \frac{1}{2(1+\xi)}
\nabla^\mu h -  \frac{1}{2(1+\xi)} \frac{Z_5}{Z_6} \nabla^\mu
\varphi, \nn \\
\wh{C}_{\mu\nu} &=& (\xi g_{\mu\nu} \Box + \nabla_\mu \nabla_\nu -
\xi R_{\mu\nu} ) Z_6. \nn
\eea
The one-loop effective action is given by the standard expression
\beq
\Gamma_{div} = \frac{i}{2}  \tr \log \wh{H}- i  \tr \log \wh{M} +
\frac{i}{2}  \tr \log \wh{C},
\label{10}
\eeq
where
\[ \wh{H}_{ij} = \left( S^{(2)} + \delta  S^{(2)} + S_{GF}
\right)_{ij}, \ \ \ \ \wh{M}_{\mu\nu} = \wh{C}_{\mu\lambda}
\frac{\delta \chi^\lambda}{\delta u^{\nu}}, \]
where $u^\nu$ are the gauge transformation parameters.

It is interesting to note that the divergent part of the last term
in (\ref{10}) is known to be a $\lambda$-dependent surface term
$\lambda = -1/(1+\xi)$, which is well defined only for $\lambda
>-
1$, i.e. $\xi >0$. Since we are not interested in surface
divergences, this term will not be important for our purposes.
The calculation of the divergences being extremely tedious, we
restrict ourselves to the particular case $\Phi=$ const., while
arguing that $\Gamma_{div}$ is off-shell identical to that of
the less technical case of the conformal gauge. For  $\Phi=$ const.
the minimal quartic operator (\ref{5}) reads (only its non-zero
components are written)
\bea
&& \wh{K}_{\varphi\varphi} = 2Z_3 - \frac{Z_5^2}{2(1+ \xi) Z_6}, \
\ \ \wh{K}_{\varphi h} =\wh{K}_{h \varphi } =  - \frac{\xi
Z_5}{2(1+ \xi)}, \ \ \ \wh{K}_{h h} =   \frac{\xi Z_6}{2(1+ \xi)},
\nn \\
&& \wh{K}_{\bar{h}_{\rho\sigma} \bar{h}_{\alpha\beta}} = - \xi
Z_6 P^{\rho\sigma,\alpha\beta}, \ \ \  P^{\rho\sigma,\alpha\beta}
\equiv g^{(\rho \alpha} g^{\sigma\beta)} - \frac{1}{2}
g^{\rho\sigma} g^{\alpha\beta}, \nn \\
&& \wh{M}^{\mu\nu}_{\varphi\varphi} = -2C_1 g^{\mu\nu} + (Z_3 -2Z_4
+2{Z'}_5 ) R g^{\mu\nu},
 \nn \\
&& \wh{M}^{\mu\nu}_{\varphi h} = \wh{M}^{\mu\nu}_{h \varphi} = -
\frac{1}{2} {C'}_2  g^{\mu\nu} - \frac{1}{2} (Z_5+2{Z'}_6)  R
g^{\mu\nu},
\ \ \ \wh{M}^{\mu\nu}_{hh} = Z_6 R g^{\mu\nu},
 \nn \\
&& \wh{M}^{\mu\nu}_{\bar{h}_{\rho\sigma} \bar{h}_{\alpha\beta}}
=\left[ \frac{C_2}{2} + (1+\xi ) Z_6 R \right]
P^{\rho\sigma,\alpha\beta} g^{\mu\nu}+ \left[ (4\xi -2) Z_6 R - C_2
\right]  P^{\rho\sigma,\mu\lambda}  P^{\alpha\beta,\nu\kappa}
g_{\kappa\lambda}.
\eea
Now, there is a simple way to see how the $\bar{h}\bar{h}$ sector
decouples, namely to set $\xi \rightarrow \infty$, for instance,
and indeed the explicit calculation shows that the
$\bar{h}\bar{h}$ terms only give $\xi$-independent total
derivatives.

The ghost operator has the form
\beq
\wh{M}^{\mu\nu} = \xi \Box^2 g_{\mu\nu} + R \nabla_\mu \nabla_\nu.
\eeq
Discarding $\xi$-independent surface terms, one easily gets the
following contribution to (\ref{10})
\beq
-\frac{1}{2\epsilon}  \int d^2x \, \sqrt{g}\,  \left[ \frac{2}{\xi}
R
+ \cdots \right].
\eeq
This is clearly a surface term as well, but we have kept it to
demonstrate that it cancels out the corresponding term stemming
from $\tr \log \wh{H}$. Performing some algebra (and using Eq.
(\ref{7})), we obtain
\bea
\Gamma_{div} [\Phi = const]& =& -\frac{1}{2\epsilon}  \int d^2x \,
\sqrt{g}\,  \left[ \frac{1}{\Delta} (4C_1 Z_6 + 2{C'}_2Z_5 )
\right.
\nn \\
&& + \left. \frac{2}{\Delta} ( Z_5^2 -5Z_3Z_6 +2Z_4Z_6 +2Z_5{Z'}_6
-2Z_6{Z'}_5 ) R \right].
\label{14}
\eea
{\it All} the $\xi$-dependent terms including the surface ones have
cancelled exactly. This is a pleasant surprise since, generally
speaking, we would have expected that a renormalization of the
metric would be needed in order to eliminate $\xi$ from the effective
action (as
in the theory (\ref{1}), see \cite{5}). We will see below that the
result of the calculation in the conformal gauge at $\Phi=$  const.
exactly coincides with Eq. (\ref{14}).

Now, working in the conformal gauge, we split the fields according
to
\beq
\Phi  \longrightarrow \Phi + \varphi, \ \ \ \ \ g_{\mu\nu}
\longrightarrow e^{2\sigma}  g_{\mu\nu}.
\eeq
Under the conformal transformation, our basic action (\ref{3})
becomes
\bea
S &=& - \int d^2x \, \sqrt{g}\, e^{-2\sigma} \left\{ Z_1(\Phi)
g^{\mu\nu} g^{\alpha
\beta} \partial_\mu \Phi  \partial_\nu \Phi  \partial_\alpha \Phi
\partial_\beta \Phi +  Z_2(\Phi) g^{\mu\nu} g^{\alpha \beta}
\partial_\mu \Phi  \partial_\alpha \Phi  \nabla_\nu  \partial_\beta
\Phi \right. \nn \\ & -& Z_2(\Phi) g^{\mu\nu} g^{\alpha \beta}
\partial_\mu \Phi  \partial_\alpha \Phi  \partial_\nu \Phi
\partial_\beta \sigma +
 Z_3(\Phi) g^{\mu\nu} g^{\alpha \beta} \nabla_\mu
\partial_\alpha \Phi  \nabla_\nu  \partial_\beta \Phi + 2 Z_3(\Phi)
g^{\mu\nu} \Box \Phi \partial_\mu \Phi \partial_\nu \sigma \nn \\
&-& 4 Z_3(\Phi) g^{\mu\nu} g^{\alpha \beta} \nabla_\mu
\partial_\alpha \Phi  \partial_\nu \Phi \partial_\beta \sigma+ 2
 Z_3(\Phi) g^{\mu\nu} g^{\alpha \beta} \partial_\mu \Phi
\partial_\nu \Phi \partial_\alpha \sigma \partial_\beta \sigma \nn
\\ &+& Z_4(\Phi)
R g^{\mu\nu}  \partial_\mu \Phi  \partial_\nu \Phi - 2Z_4(\Phi)
 g^{\mu\nu}  \partial_\mu \Phi  \partial_\nu \Phi \Box \sigma  +
Z_5(\Phi) R \Box  \Phi  \nn \\ &-& 2Z_5(\Phi) \Box \sigma \Box
\Phi  + Z_6(\Phi) R^2 -4Z_6(\Phi) R \Box \sigma + 4Z_6(\Phi) \Box
\sigma \Box \sigma\nn \\
&+& \left. e^{2\sigma} \left[  C_1(\Phi) g^{\mu\nu}  \partial_\mu
\Phi  \partial_\nu \Phi  +   C_2(\Phi) R -2  C_2(\Phi)\Box \sigma
\right] + V(\Phi) e^{4\sigma} \right\}.
\label{16}
\eea
The corresponding ghost contribution to the divergences are just
the surface terms and hence they will be dropped. Thus, we have an
effective theory of two scalars on a curved background.

Expanding in powers of the quantum fields $\Phi^i \equiv \{
\varphi, \sigma\}$, we get
\bea
&&\wh{K}_{ij} =2 \left( \brr{cc} Z_3 & -Z_5 \\ -Z_5 & 4Z_6 \err
\right), \ \ \ \ \wt{L}_{ij}^{\mu\nu\lambda} = \wt{A}^\rho_{ij}
g_\rho^{(\lambda} g^{\mu\nu)}, \nn \\
&&  \wh{A}_{\varphi\varphi}^\lambda = 4({Z'}_3 -Z_2)
\partial^\lambda \Phi , \ \ \    \wh{A}_{1\varphi\sigma}^\lambda =
4(2Z_4 -{Z'}_5) \partial^\lambda \Phi , \ \ \
\wh{A}_{\sigma\varphi}^\lambda = 4(Z_3 -{Z'}_5) \partial^\lambda
\Phi , \nn \\
&& \wh{A}_{\sigma\sigma}^\lambda = 16{Z'}_6 \partial^\lambda \Phi
, \ \ \ \ \ \wh{M}_{ij} = g_{\mu\nu}\wh{M}_{ij}^{\mu\nu}, \nn \\
&& \wh{M}_{\varphi\varphi} = -4C_1 +2(Z_3-2Z_4+2 {Z'}_5)R + 2(
{Z''}_3-8Z_1 -{Z'}_2) \partial^\lambda \Phi \partial_\lambda \Phi
+6({Z'}_3-Z_2) \Box \Phi, \nn \\
&& \wh{M}_{\varphi\sigma} =\wh{M}_{\sigma\varphi} = -4{C'}_2 -
4(Z_5+2 {Z'}_6)R + 4(2Z_4-Z_3-2{Z'}_5) \Box \Phi \nn \\ && \ \ \ \
\ \ \ \ \ \ \
\ \ \ + 2(Z_2
+2{Z'}_4 -2{Z''}_5) \partial^\lambda \Phi \partial_\lambda \Phi,
\nn \\
&& \wh{M}_{\sigma\sigma} = 32Z_6R + 16(Z_5+{Z'}_6) \Box \Phi +
8(2Z_4 -Z_3+2{Z''}_6) \partial^\lambda \Phi \partial_\lambda \Phi.
\eea
Following the procedure (\ref{6})--(\ref{7}), after some algebra
the one-loop effective action can be found to be (having dropped
the surface terms)
\bea
\Gamma_{div} & =& -\frac{1}{2\epsilon}  \int d^2x \,
\sqrt{g}\,  \left[ \frac{1}{\Delta} (4C_1 Z_6 + 2{C'}_2Z_5 ) +
\frac{2}{\Delta} (Z_5^2-5Z_3Z_6+2Z_4Z_6+2Z_5 {Z'}_6 -2Z_6{Z'}_5 )
R \right. \nn \\
&&+ \frac{1}{\Delta} \left( \frac{3}{2} Z_3^2 +16Z_1Z_6 -Z_2Z_5 -
2Z_3Z_4 -2Z_4^2 +2Z_5{Z'}_3+2Z_5{Z'}_4 -6Z_6{Z'}_2 \right. \nn \\
&& \left. \left. +2Z_6{Z''}_3 \right) \partial^\lambda \Phi
\partial_\lambda \Phi
+ \left(\frac{2 ( Z_2Z_6 -Z_5{Z'}_5 +Z_6{Z'}_3)}{\Delta}
\right)' \partial^\lambda \Phi \partial_\lambda \Phi \right] \nn \\
&& \equiv  -\frac{1}{2\epsilon}  \int d^2x \,
\sqrt{g}\,  \left[ A_1(\Phi) +  A_2(\Phi) R+  A_3(\Phi)
\partial^\lambda \Phi \partial_\lambda \Phi \right].
\label{18}
\eea
As can be easily seen, for $\Phi =$ const. Eq. (\ref{18}) coincides
with the covariant gauge effective action (\ref{14}). Notice also
that higher-derivative divergences do not appear.

The theory under discussion is one-loop multiplicatively
renormalizable in the usual sense if the following conditions are
fulfilled
\beq
A_1(\Phi) = a_1 V(\Phi), \ \ \  A_2(\Phi) = a_2 C_2(\Phi), \ \ \
A_3(\Phi) = a_3 C_1(\Phi),
\label{19}
\eeq
where $a_1$, $a_2$ and $a_3$ are arbitrary constants. Many different
sets of dilatonic functions in (\ref{3}) satisfy the conditions
(\ref{19}). A simple example of a multiplicatively renormalizable
theory is given by
\beq
Z_i=e^\Phi, \ \ i=1, \ldots, 6, \ \ \ C_1 = const, \ \ \ C_2 =
const, \ \ \ V = const,
\eeq
or the even more immediate one $Z_i=1$, $i=1, \ldots, 6$, $C_1
(\Phi)$, $C_2 (\Phi)$ and $V(\Phi)$ being arbitrary. Some other
families of solutions can  also be given explicitly.

It is easy to write the generalized renormalization group
equations
for our theory. In particular, all generalized $\beta$-functions
corresponding to the higher-derivative terms are zero, and we have
a large freedom because all the functions $Z_i$ are free parameters
of the theory, in the generalized renormalization group.

To be noted also is the fact that one cannot obtain the one-loop
renormalization of low-derivative 2D dilaton gravity (\ref{1}) as
a particular case of the theory (\ref{3}). For that purpose it is
necessary to put the $Z_i$ in (\ref{3}) equal to zero, but such a
restriction contradicts condition (\ref{4}). The reason is that
higher-derivative terms give essential contributions to the
renormalization of the low-derivative terms (but not to their own
renormalization!). This fact completely changes the structure of
renormalization, as compared with that of low-derivative dilaton
gravity \cite{3}.
\bs

\noindent 3. {\bf $\sigma$-model interpretation}.
The use of the conformal gauge indicates the possibility to
interprete the theory under discussion as a certain $D=2$
$\sigma$-model  with higher derivatives. Indeed, choosing the
conformal gauge $g_{\mu\nu} \rightarrow e^{2\sigma} g_{\mu\nu}$,
one can represent the theory (\ref{16}) as
\bea
S&=&  \int d^2x \, \sqrt{g}\,  \left[ T(X) + R\psi (X) + G_{ab} (X)
g^{\mu\nu} \partial_\mu X^a \partial_\nu X^b + C(X)R^2 + R \Box X^a
U_a(X) \right. \nn \\
&&+ R g^{\mu\nu} \partial_\mu X^a \partial_\nu X^b  W^{(1)}_{ab}
(X) + g^{\mu\nu} g^{\alpha\beta} \partial_\mu X^a \partial_\nu X^b
\partial_\alpha X^c \partial_\beta X^d F^{(1)}_{abcd} (X) \nn \\
&&+  g^{\mu\nu} \Box X^a \partial_\mu X^b \partial_\nu X^c
T^{(1)}_{abc} (X) + g^{\mu\nu} g^{\alpha\beta} \nabla_\mu
\partial_\alpha X^a \partial_\nu X^b  \partial_\beta X^c
T^{(2)}_{abc} (X) \nn \\
&& + \left. \Box X^a \Box X^b M_{ab}^{(1)} (X) + g^{\mu\nu}
g^{\alpha\beta} \nabla_\mu \partial_\alpha X^a \nabla_\nu
\partial_\beta X^b M^{(2)}_{ab} (X) \right],
\label{21}
\eea
where \[ X^a = (\sigma, \Phi), \ \ \ T(X)= - V(\Phi) e^{2\sigma},
\ \ \ \psi (X) = -C_2(\Phi), \ \ \ C(X)= -e^{-2 \sigma} Z_6(\Phi),
\] \[ G_{ab} = - \left( \brr{cc} 0 & {C'}_2 (\Phi) \\ {C'}_2 (\Phi)
& C_1(\Phi ) \err \right) , \ \ (U_\Phi, U_\sigma) = - e^{-2\sigma}
(Z_5(\Phi), -4 Z_6 (\Phi)), \] \[ W^{(1)}_{ab} =- e^{-2\sigma} \left(
\brr{cc} 0 & 0 \\ 0 & Z_4(\Phi ) \err \right), \]
and the non-zero components of the remaining tensors are
\bea
&& F^{(1)}_{\Phi\Phi\Phi\Phi} = -e^{-2 \sigma} Z_1(\Phi), \ \ \
F^{(1)}_{\Phi\Phi\Phi\sigma} = e^{-2 \sigma} Z_2(\Phi), \ \ \
F^{(1)}_{\Phi\Phi\sigma\sigma} =-2 e^{-2 \sigma} Z_3(\Phi), \nn \\
&& T^{(1)}_{\Phi\Phi\sigma} = -2e^{-2 \sigma} Z_3(\Phi), \ \
T^{(1)}_{\sigma\Phi\Phi} =2 e^{-2 \sigma} Z_4(\Phi), \ \
T^{(2)}_{\Phi\Phi\Phi} =- e^{-2 \sigma} Z_2(\Phi), \ \
T^{(2)}_{\Phi\Phi\sigma} =4 e^{-2 \sigma} Z_3(\Phi), \nn \\
&& M^{(1)}_{\sigma\sigma} = -4e^{-2 \sigma} Z_6(\Phi), \ \ \
M^{(1)}_{\sigma\Phi} =2 e^{-2 \sigma} Z_5(\Phi), \ \ \
M^{(2)}_{\Phi\Phi} =- e^{-2 \sigma} Z_3(\Phi).
\eea
Thus, we have arrived to the particular case of $D=2$ higher-derivative
$\sigma$-model considered in Ref. \cite{1}. The field
$M^{(2)}_{ab}$ is of Stueckelberg type, and can be gauged away by
integrating by parts (this changes some other higher-derivative
terms in (\ref{21})). The background field equations in the linear
approximation (which describe the first massive level of the
corresponding string) can be easily taken from \cite{1} for our
specific values of the functions under discussion.
\bs

\noindent 4. {\bf Dilatonic solutions}.
To simplify the analysis a bit, we start from the following version
of action (\ref{3})
\bea
S &=& - \int d^2x \, \sqrt{g}\,  \left[ Z_1 g^{\mu\nu}
g^{\alpha
\beta} \partial_\mu \Phi  \partial_\nu \Phi  \partial_\alpha \Phi
\partial_\beta \Phi +
Z_5 R \Box  \Phi  + Z_6 R^2 \right. \nn \\
&&+ \left.  C_1(\Phi) g^{\mu\nu}  \partial_\mu \Phi  \partial_\nu
\Phi  + V(\Phi) \right],
\label{da}
\eea
where $Z_1$, $Z_5$ and $Z_6$ are constants, and always $Z_5 \neq 0$
and  $Z_6 \neq 0$. The corresponding equations of motion are
\bea
\frac{\delta S}{\delta \Phi} &=& -4Z_1 \nabla_\alpha \left(
g^{\mu\nu} g^{\alpha
\beta} \partial_\mu \Phi  \partial_\nu \Phi  \partial_\beta \Phi
\right)  + Z_5 \Box R  + V'(\Phi)  \nn \\
&&+   {C'}_1(\Phi) g^{\mu\nu}  \partial_\mu \Phi  \partial_\nu
\Phi - 2\nabla_\nu \left[ C_1(\Phi) g^{\mu\nu}  \partial_\mu \Phi
\right] =0, \nn \\
\frac{1}{2} g^{\mu\nu} \frac{\delta S}{\delta g_{\mu\nu}} &=& -Z_1
 g^{\mu\nu} g^{\alpha
\beta} \partial_\mu \Phi  \partial_\nu \Phi  \partial_\alpha
\Phi\partial_\beta \Phi  + Z_5 \Box^2 \Phi -Z_5 R \Box \Phi
\nn \\
&& - Z_6 R^2 + 2Z_6 \Box R + V(\Phi ) =0.
\eea
We will deal with black hole type metrics of the ordinary kind (using
the gauge \cite{ln}) \beq
ds^2 = -g(r) dt^2 +g^{-1} (r) dr^2 \longrightarrow -dt^2 +dr^2, \
\ r \rightarrow \infty.
\eeq
Taking into account the $t$-independence of $\Phi$ (and meaning
now $(\ )' \equiv  \partial_r$), we obtain
\bea
&& -4Z_1 \left[ g^2 (\Phi')^3 \right]' + Z_5 \left( gg'''\right)'
+ V'(\Phi )/ \Phi' - \left[C_1 (\Phi) g \Phi' \right]' -  C_1
(\Phi) \left( g \Phi' \right)' =0, \nn \\
&& -Z_1 g^2 (\Phi')^4 + Z_5 \left[ g (g\Phi')''\right]' - Z_5 g''
(g\Phi')'-Z_6 (g'')^2 + 2Z_6 (gg'')' +V (\Phi) =0.
\label{em2}
\eea
{}From previous analysis of standard dilatonic gravity, we can expect to
find
solutions of these quite involved differential equations (\ref{em2})
for string potentials of the form
\beq
V \sim \Lambda + e^{\lambda \Phi (r)}.
\label{pot}
\eeq
The solutions are of the kind \beq
\Phi (r) = a+br,
\eeq
 being $\Lambda$, $\lambda$, $a$ and $b$ some constants. This is
indeed the case. By expanding $g(r)$ in series
\beq
g = c + \frac{\alpha}{r} - \frac{\alpha_1}{r^2} + \cdots, \ \ \
c=g(\infty),
\label{gex}
\eeq
it is not difficult to see that indeed solutions of (\ref{em2}) are
found in the following two cases.
\medskip

\noindent{\it (a) Case $Z_1 \neq 0, \ C_1 =$ const}. The equations of
motion reduce to (with $\Phi = a+br$)
\bea
&& -4Z_1 b^4  g^2  + Z_5 b g g''' + V(\Phi ) - 2C_1 b^2 g = k, \nn
\\
&& -Z_1 b^4 g^2  + Z_5 b g g''' -Z_6 (g'')^2 + 2Z_6 (gg'')' +V
(\Phi) =0,
\label{em3}
\eea
where $k$ is an arbitrary constant (of integration).
A solution is found for arbitrary $\Lambda$, with
\beq
b^2 = - \frac{C_1}{3Z_1 c},
\eeq
(in other words, satisfying $\Phi'(r)^2 \simeq -C_1 /(3Z_1
g(1/r))$), with the only restriction that $\alpha =0$, i.e., that
\beq
\left. \frac{dg}{dr^{-1}} \right|_{r^{-1} =0} =0.
\eeq
This solution is obtained by substituting the expansion (\ref{gex})
into the equations of motion (\ref{em2}), and it is exact up to
terms of order ${\cal O} (r^{-4})$. That all terms up to this order
can be matched for both equations with so few requeriments is not
trivial at all, as can be seen immediately by comparing with different
ans\"{a}tze.
\medskip

\noindent{\it (b) Case} $Z_1 = 0, \ C_1 (\Phi ) \sim V (\Phi )$. This is
also an interesting situation and the result is very similar to the
previous one. The equations of motion can now be written as (again
with $\Phi = a+br$)
\bea
&&  Z_5 b \left( gg'''\right)' + V'(\Phi ) -2 b^2 \left[C_1 (\Phi) g
\right]' +  {C'}_1 (\Phi) b^2 g =0, \nn \\
&&  Z_5 b g g''' -Z_6 (g'')^2 + 2Z_6 (gg'')' +V (\Phi) =0.
\label{em4}
\eea
As before, a solution which is exact up to order  ${\cal O} (r^{-
4})$ is obtained for
\beq
b^2 = \frac{V'(\infty)}{g(\infty) {C'}_1 (\infty)}
\eeq
(this is a non-zero constant), for a potential of the form
(\ref{pot}). (Actually, to match the terms up to order four in
$r^{-1}$, also potentials e.g. of the form $V=V_0 \left( 1- e^{-v/r^4}
\right)$ would do, but these particular forms are too connected with
the approximation one is working at).
\medskip

In contrast, it is also easy to check that an ansatz of the type
$\Phi = a + \log (r-r_0)$ does {\it not} lead to any solution in the
first case, unless $g=$ const., a trivial situation. In the second
case, one could say that a solution is obtained (in principle) to
order  ${\cal O} (r^{-4})$, since all terms do vanish to this
order. However, it is not really meaningful as an approximation to
a series solution, since already the first non-vanishing terms
cannot be compensated in any way.
\bs

\noindent 5. {\bf Concluding remarks}.
Of course many questions are still left for future investigation of
such higher-derivative $D=2$ $\sigma$-model. Some of them have been
listed in Ref. \cite{1}. One of particular importance for us
concerns the relations between the $\beta$-functions corresponding
to the couplings in (\ref{21}) or, more precisely, which of these
$\beta$-functions are independent? The answer for the case of the
standard $\sigma$-model (first three terms in (\ref{21})) is well
known (see for example \cite{7}), but not for the full
higher-derivative
$\sigma$-model
(\ref{21}). As we could see from the calculations in the previous
section, only $T(X)$, $\psi (X)$ and $G_{ab} (X)$ are getting
renormalized in the one-loop approximation, and only through
higher-derivative terms (or in the case of $T(X)$ through a mixture
of higher-derivative and lower-derivative terms).

The other interesting question which remains open concerns the
interpretation of the condition of the vanishing of the
$\beta$-functions in the model under discussion. However, all these
questions should be first understood for the case of the general
higher-derivative $\sigma$-model \cite{1}.

Note, finally, that the model discussed in this work provides a big
arena for the study of 2D black holes and their properties, like Hawking
radiation \cite{9}, etc., and surely deserves further study.

\vspace{5mm}

\noindent{\large \bf Acknowledgments}

Thanks are given to Iosef Buchbinder and to Hugh Osborn for helpful
discussions. This work has been  supported by DGICYT (Spain) and by
CIRIT (Generalitat de Catalunya).

\end{document}